\documentclass[lettersize,journal]{IEEEtran}
\IEEEoverridecommandlockouts
\usepackage{booktabs}                 % 提供 \toprule, \midrule, \bottomrule 等命令
\usepackage{multirow}                 % 如果需要多行合并单元格
\usepackage{array}                    % 增强版表格功能
\usepackage{caption}                  % 对表格或图片标题进行控制
\usepackage{float}                    % 提供 [H] 选项以固定位置
\usepackage{cite}
\usepackage{adjustbox}
\usepackage{makecell}
\usepackage{amsmath,amssymb,amsfonts}
\usepackage{graphicx}
\usepackage{textcomp}
\usepackage{xcolor}
\usepackage{adjustbox}
\usepackage{tabularx} % for 'tabularx' env. and 'X' col. type
\usepackage{ragged2e} % for \RaggedRight macro
\usepackage{siunitx}
\usepackage{algorithm}
\usepackage{algorithmicx}  % Use algorithmicx instead of algorithmic
\usepackage{algpseudocode} % For better pseudocode formatting
\usepackage{amsmath}
\usepackage{amssymb}

\usepackage{threeparttable}
\usepackage{booktabs}
\usepackage{hyperref}
\usepackage{graphicx}
\usepackage{subcaption}
\def\BibTeX{{\rm B\kern-.05em{\sc i\kern-.025em b}\kern-.08em
    T\kern-.1667em\lower.7ex\hbox{E}\kern-.125emX}}
\begin{document}

\title{URL2Graph++: Unified Semantic-Structural-Character Learning for Malicious URL Detection\\

}

\author{
Ye Tian$^{\dagger}$,
Yifan Jia$^{\dagger}$,  
Yanbin Wang\textsuperscript{*}, 
Jianguo Sun\textsuperscript{*}, 
Zhiquan Liu,
Xiaowen Ling
\thanks{$^{\dagger}$ Ye Tian and Yifan Jia contributed equally to this work.}%
\thanks{\textsuperscript{*} Corresponding authors: Yanbin Wang and Jianguo Sun.}%
\thanks{Yifan Jia is with Yantai Research Institute, Harbin Engineering University, Yantai, China (email: jiayf@hrbeu.edu.cn).}%
\thanks{Ye Tian and Jianguo Sun are with Hangzhou Research Institute, Xidian University, Hangzhou, China (emails: tianye@xidian.edu.cn; jgsun@xidian.edu.cn).}%
\thanks{Yanbin Wang is with Shenzhen MSU-BIT University, Shenzhen, China, and School of Computer Science and Technology, Beijing Institute of Technology, Beijing, China (email: wangyanbin15@mails.ucas.ac.cn).}%
\thanks{Zhiquan Liu is with the College of Cyber Security, Jinan University, Guangzhou, China (email: zqliu@vip.qq.com).}%
\thanks{Xiaowen Ling is with Shenzhen MSU-BIT University, Shenzhen, China (email: 1120220517@smbu.edu.cn).}%
}

\maketitle

\begin{abstract}
Malicious URL detection remains a major challenge in cybersecurity, primarily due to two factors: (1) the exponential growth of the Internet has led to an immense diversity of URLs, making generalized detection increasingly difficult; and (2) attackers are increasingly employing sophisticated obfuscation techniques to evade detection. We advocate that addressing these challenges fundamentally requires: (1) obtaining semantic understanding to improve generalization across vast and diverse URL sets, and (2) accurately modeling contextual relationships within the structural composition of URLs. In this paper, we propose a novel malicious URL detection method combining multi-granularity graph learning with semantic embedding to jointly capture semantic, character-level, and structural features for robust URL analysis. To model internal dependencies within URLs, we first construct dual-granularity URL graphs at both subword and character levels, where nodes represent URL tokens/characters and edges encode co-occurrence relationships. To obtain fine-grained embeddings, we initialize node representations using a character-level convolutional network. The two graphs are then processed through jointly trained Graph Convolutional Networks (GCNs) to learn consistent graph-level representations, enabling the model to capture complementary structural features that reflect co-occurrence patterns and character-level dependencies. Furthermore, we employ BERT to derive semantic representations of URLs for semantically aware understanding. Finally, we introduce a gated dynamic fusion network to combine the semantically enriched BERT representations with the jointly optimized graph vectors, further enhancing detection performance. We extensively evaluate our method across multiple challenging dimensions: real-world data distributions, generalization, character confusion, and short URLs—covering key practical challenges. Results show our method exceeds SOTA performance, including against large language models. Our source code is available at: \url{https://github.com/lincozz/URL2Graphplusplus}.
\end{abstract}


\section{Introduction}

Malicious URLs are deceptive web links crafted to facilitate phishing, fraud, malware distribution, and command-and-control, often by impersonating trusted brands and exploiting hyperlink presentation or redirection \cite{sabir2022reliability,bitaab2023beyond}. Such links jeopardize individuals and organizations by enabling credential theft, privacy leakage, and service disruption \cite{cao2025phishagent,liu2022inferring,li2025phishintel,li2024knowphish}. Recent industry and law‑enforcement reports indicate sustained growth in both phishing volume and financially consequential incidents \cite{liu2024less}. According to the FBI's 2024 Internet Crime Complaint Center (IC3) report, cybercrime resulted in a record \$16.6 billion in reported losses—representing a 33\% increase over the previous year—while phishing and spoofing remained the most frequently reported types of crime
\cite{fbi_ic3_2024}. These trends underscore the necessity of developing advanced detection approaches that achieve both high accuracy and operational robustness in adversarial environments.

Conventional defenses—blacklists, heuristic filters, and rule-based systems —remain necessary but exhibit delayed coverage and brittleness to novel or obfuscated URLs \cite{sahoo2017malicious,liang2021robust,lin2021phishpedia,aljofey2022effective}. Machine learning has improved recall by leveraging lexical patterns such as length, token frequencies, and character n-grams \cite{yuan2018url2vec}, and recent deep architectures further exploit subword semantics and contextual cues. Yet adversaries increasingly deploy short links and semantic camouflage, use subdomain nesting and path rewriting, and apply homoglyph substitution to evade sequential detectors. In practice, effective systems must sustain performance under distribution shift, operate at low false-positive rates, generalize across heterogeneous data sources, and integrate both high-level semantic cues and fine-grained morphological signals.

Current approaches to malicious URL detection rely on sequential representations from URLs using CNNs, RNNs, or Transformers. However, these methods fundamentally ignore non-sequential relational patterns that define advanced attacks, facing three limitations that hinder their effectiveness against evolving threats:
\begin{itemize}
    \item Non-local pattern blindness — CNNs and RNNs inherently focus on local sequential patterns, failing to capture critical long-range dependencies between semantically related but positionally distant tokens (e.g., the suspicious association between "account" and "verify" in "paypal.com/account/id235/secure/verify.php" despite 4 intermediate tokens).
    \item Structural ignorance — existing models process URLs as linear strings, disregarding the inherent graph-like structure of URLs (including host-path hierarchies and query parameter relationships), which attackers systematically exploit through techniques like subdomain nesting ("login.mail.service.paypal.com.confirm@phishing.com") or path obfuscation.
    \item Character-level myopia — while some works employ character-CNNs, they only detect local n-gram patterns without modeling systematic character-level attack strategies such as homoglyph repetition ("g00gle.com" with consecutive zero substitutions), deliberate misspellings ("faceb00k-login") \cite{lain2025url}, or abnormal Unicode distributions. These limitations collectively create measurable vulnerabilities. 
\end{itemize}

To address these limitations, we propose URL2Graph++, a multi-granularity learning framework that unifies semantic, structural, and character-level signals through four key components: (1) dual-feature encoding where BERT extracts contextual subword embeddings while CharCNN captures character-level anomalies; (2) dual-granularity graph construction that builds both subword-level graphs (with nodes initialized by combining token embeddings and CharCNN features to capture semantic-character interactions) and character-level graphs (with nodes initialized by CharCNN outputs for morphological pattern analysis); (3) dual-graph learning with jointly trained GNNs — the subword GNN identifies suspicious token relationships while the character GNN detects obfuscation patterns, optimized through a shared loss function to align cross-granularity features; and (4) adaptive fusion through a gating network that dynamically combines BERT embeddings, subword-GNN outputs, and character-GNN outputs to effectively handle complex URL attacks ranging from semantic phishing to character-level obfuscation.

The key contributions of this work are:

\begin{itemize}
    \item We propose the first cascaded fusion architecture that progressively combines three complementary information: (a) deep URL semantic , (b) character-level anomaly, and (3) global dependencies patterns - achieving robust URL threat assessment.
    \item We propose the first method to model URLs as graphs, introducing a dual-grained graph method where subword-level and character-level graph representations are jointly trained to capturing (1) semantic-functional co-occurrence patterns (e.g., malicious token sequences like /admin/login.php), (2) anomalous character-level dependencies (e.g., random strings or homoglyphs).
    \item Our work establishes new SOTA performance in URL analysis, demonstrating: (a) 9\% higher detection accuracy for malicious short URLs than previous best methods, (b) consistent outperformance over the most advanced baselines, including both URL-optimized BERT variants and general-purpose large language models (LLMs), and (c) robustness across real-world data distributions, generalization tests, and character obfuscation scenarios.
\end{itemize}

\section{Related Work}
Malicious URL detection has been extensively studied at the intersection of cybersecurity and machine learning. As attackers increasingly exploit semantic obfuscation, structural manipulation, and character-level perturbations, the community has progressed from manual feature engineering to end-to-end neural architectures that model URLs at multiple granularities. In what follows, we review four lines of research most relevant to our work: traditional machine learning with handcrafted features, deep sequential models and approaches based on pre-trained language models.

\subsection{Traditional Machine Learning with Handcrafted Features}
Early work cast malicious URL detection as a supervised learning problem built on handcrafted representations that aggregate lexical statistics, token and n-gram frequencies, domain- and path-level cues, and character distributions \cite{zhang2011textual,ma2011learning,mamun2016detecting,jain2017phishing}. Classical classifiers—including support vector machines, logistic regression, random forests, and ensemble methods—were then trained on these feature sets \cite{almomani2018fast,chiramdasu2021malicious,rupa2021machine,ullah2022malware,li2019stacking}. Although such pipelines are computationally efficient and offer clear interpretability, their reliance on fixed feature templates reduces resilience to polymorphic patterns and deliberate obfuscation. Moreover, while suitable for smaller corpora, manual feature construction scales poorly and tends to overlook informative signals in semantically complex and structurally heterogeneous URLs, thereby limiting the ability to model higher-order semantic relationships and contextual dependencies.

\subsection{Deep Learning and Sequential Modeling}
With the advent of deep learning, CNN- and RNN-style models began to replace manual feature pipelines by learning representations directly from raw URLs \cite{tsai2024toward,sahingoz2024dephides}. URLNet \cite{le2018urlnet} popularized a dual-channel CNN that jointly encodes character- and word-level signals; subsequent works extend this idea with deeper convolutional blocks, attention mechanisms, and hybrid CNN–RNN backbones \cite{tajaddodianfar2020texception,wang2022tcurl,hussain2023cnn,zheng2022hdp,srinivasan2021durld,yan2020learning}. Capsule-enhanced and bidirectional recurrent variants further improve sequential feature extraction and local pattern capture \cite{huang2019phishing,wang2019bidirectional}. GramBeddings \cite{bozkir2023grambeddings} combines CNNs, LSTMs, and attention over n-gram features, reporting competitive results on large-scale corpora. While these models reduce the need for feature engineering and deliver strong performance, they treat URLs primarily as linear sequences. As a result, non-local dependencies between semantically related but distant tokens, and hierarchical relations across host–path–query components, remain only implicitly modeled, which may degrade robustness under distribution shift and heavy obfuscation.

\subsection{Pre-trained Language Models for URL Encoding}
Inspired by advances in NLP, pre-trained Transformer encoders have been adapted to malicious URL detection by tokenizing URLs and leveraging contextualized subword embeddings \cite{chang2021research,xu2021transformer,li2025urlbert}. URLTran \cite{maneriker2021urltran} systematically evaluates Transformer backbones and training regimes for phishing detection, demonstrating the feasibility of transferring language-modeling capacity to URL strings. Other efforts fine-tune BERT variants directly on URL corpora or explore lightweight Transformer designs for efficiency \cite{chang2021research,xu2021transformer,wang2023lightweight}. A line of work pre-trains domain-specific BERT models from scratch on large URL datasets \cite{wang2023large}, which mitigates domain mismatch at the expense of substantial data and computation. Despite their strong sequence modeling capabilities, PLM-based approaches still face challenges with rare subdomains, mixed encodings, and character-level perturbations that are not fully captured by subword tokenization alone. This motivates architectures that complement PLM semantics with explicit character modeling and structural priors.

Although sequential encoders and pre-trained Transformers have significantly advanced malicious URL detection, most existing studies continue to regard URLs as purely linear strings. This assumption neglects two critical aspects: long-range dependencies between distant but semantically correlated tokens, and systematic character-level manipulations that adversaries frequently exploit. Recent work has begun to explore graph-based approaches in this domain, but these efforts primarily focus on modeling inter-domain, HTML/DOM or host-level relationships, leaving the semantic and structural dependencies within individual URLs largely underexplored. Consequently, existing systems often struggle with short URLs that lack sufficient lexical cues, polymorphic variants produced by token reordering, and obfuscation techniques such as homoglyph substitution or random string injection. These limitations highlight the need for methods that can jointly capture semantic, structural, and character-level patterns in a unified framework, thereby enabling more robust and generalizable detection of malicious URLs under diverse and adversarial conditions. To overcome these limitations, we treat URLs as multi-granularity graph objects and explicitly learn token co-occurrence structures across subword and character levels, enabling more robust and generalizable detection.

\begin{figure*}[t]
    \centering
    \includegraphics[width=1\textwidth]{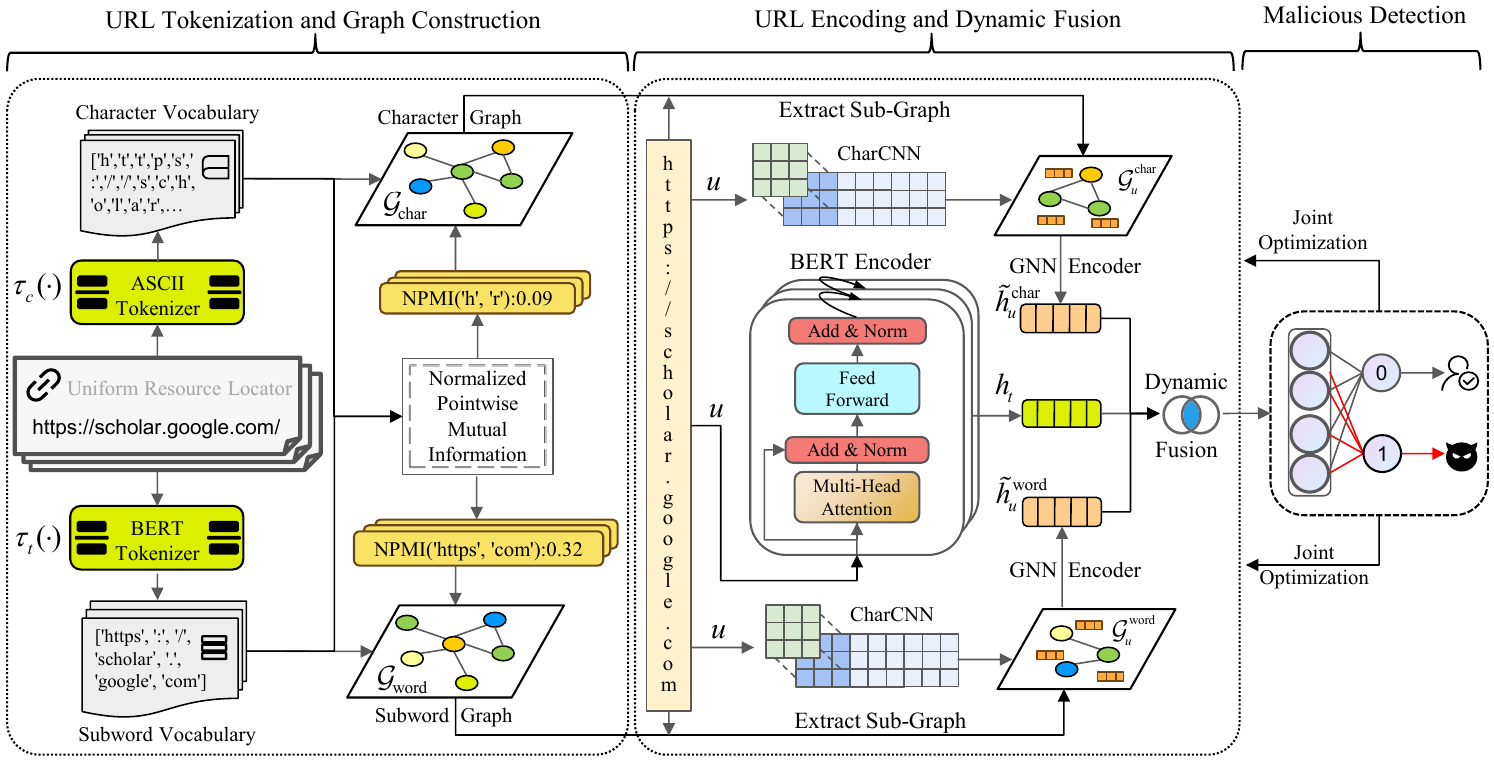}
    \caption{Architecture of the proposed URL2Graph++ framework, illustrating dual-granularity graph construction, feature encoding, and dynamic fusion.}
    \label{fig:framework}
\end{figure*}

\section{Methodology}

In this section, we present \textbf{URL2Graph++}, a multi-granularity learning framework designed to capture semantic, structural, and character-level patterns for malicious URL detection. An overview of the proposed \textit{URL2Graph++} framework is depicted in Figure~\ref{fig:framework}. Departing from purely sequential modeling approaches, our method constructs and leverages dual-grained graph representations at both subword and character levels, enabling it to capture global co-occurrence dependencies and morphological anomalies that are often exploited in obfuscated attacks.The architecture consists of three main components. First, we construct two global URL graphs based on token-level and character-level co-occurrence statistics across the dataset. These graphs encode corpus-wide relational patterns and serve as the structural basis for URL-specific subgraph sampling. Second, we apply a dual-encoder strategy in which each URL is processed by a semantic encoder and a graph encoder. The semantic encoder generates contextualized token representations, while the graph encoder models structural dependencies using subgraphs sampled from the global graphs. Finally, we introduce a gated fusion mechanism to dynamically integrate multi-level representations derived from BERT, subword-level GCN, and character-level GCN outputs. This fusion facilitates robust modeling of complex URL threats and enables joint optimization of all components in an end-to-end learning framework.

The remainder of this section details each stage of the proposed framework. We begin with the construction of global dual-grained URL graphs, which serve as the structural backbone of the model. We then describe the dual-encoder mechanism that processes individual URLs by combining sequential and graph-based encoding. Lastly, we introduce the fusion strategy that unifies multi-granularity information and enables end-to-end learning.

\subsection{Global Dual-Grained URL Graph Construction}

To capture the latent structural regularities and morphological dependencies prevalent in malicious URL patterns, we construct two global co-occurrence graphs over the entire training corpus: a \textit{subword-level graph} and a \textit{character-level graph}. These graphs serve as foundational knowledge structures that encode cross-sample token and character associations beyond individual URLs. Each graph is built based on normalized pointwise mutual information (NPMI) scores derived from corpus-wide co-occurrence statistics, allowing the model to exploit long-range dependencies and structural anomalies common in obfuscated URL attacks.

Let the training dataset be denoted as $\mathcal{D} = \{ u^{(1)}, u^{(2)}, \dots, u^{(N)} \}$, where each $u^{(i)}$ is a raw URL string. We define two tokenization functions: $\tau_t(\cdot)$ as the BERT-based subword tokenizer and $\tau_c(\cdot)$ as the character-level tokenizer that maps each URL to its ASCII character sequence. Applying these functions to the dataset yields the tokenized corpora:
\begin{equation}
\mathcal{C}_t = \bigcup_{i=1}^{N} \tau_t(u^{(i)}), \quad \mathcal{C}_c = \bigcup_{i=1}^{N} \tau_c(u^{(i)}),
\end{equation}
from which we derive the subword vocabulary $\mathcal{V}_t$ and character vocabulary $\mathcal{V}_c$ respectively as:
\begin{equation}
\mathcal{V}_t = \{w \mid w \in \mathcal{C}_t\}, \quad \mathcal{V}_c = \{c \mid c \in \mathcal{C}_c\}.
\end{equation}

\textbf{Subword-Level Graph Construction.} We begin by tokenizing each URL in the corpus using a pre-trained BERT tokenizer, which decomposes each URL string into a sequence of subword tokens from a fixed vocabulary. Let $\mathcal{V}_t = \{w_1, w_2, \dots, w_{|\mathcal{V}_t|}\}$ denote the subword vocabulary extracted from the training set. For each URL, we treat the entire tokenized sequence as a context window and collect co-occurrence counts for all unordered token pairs appearing within the same URL. We define the empirical joint probability $p(w_i, w_j)$ as the frequency with which subwords $w_i$ and $w_j$ co-occur, divided by the total number of observed subword pairs across all URLs. Similarly, the marginal probabilities $p(w_i)$ and $p(w_j)$ are computed based on unigram frequencies. The NPMI score for a pair of subwords $(w_i, w_j)$ is defined as:

\begin{equation}
\text{NPMI}( w_i, w_j ) = \frac{ \log \frac{ p( w_i, w_j ) }{ p( w_i ) p( w_j ) } }{ - \log p( w_i, w_j ) }
\end{equation}

We construct the subword-level graph $\mathcal{G}_\text{word} = (\mathcal{V}_t, \mathcal{E}_\text{word})$ by introducing an undirected edge $(w_i, w_j) \in \mathcal{E}_\text{word}$ if $\text{NPMI}(w_i, w_j) > \theta_t$, where $\theta_t$ is a subword-level threshold chosen to filter out statistically insignificant co-occurrences. The resulting graph encodes high-confidence functional and semantic associations between subword units, which are often disrupted or manipulated in adversarial URL structures.

\textbf{Character-Level Graph Construction.} Analogously, we construct a character-level graph by decomposing each URL into its constituent ASCII characters. Let $\mathcal{V}_c = \{c_1, c_2, \dots, c_{|\mathcal{V}_c|}\}$ denote the set of distinct characters observed in the corpus. Each URL forms a character sequence, and character co-occurrence counts are collected within each full URL string. The NPMI score for each character pair $(c_i, c_j)$ is computed in the same manner as the subword case, using character-level co-occurrence and marginal probabilities. The character-level graph $\mathcal{G}_\text{char} = (\mathcal{V}_c, \mathcal{E}_\text{char})$ is constructed by connecting character nodes with an edge if their NPMI score exceeds a threshold $\theta_c$.

These two global graphs capture complementary relational patterns. The subword graph encodes semantically meaningful token co-occurrences and reflects common malicious token structures, such as login-verification or credential-submission paths. The character-level graph, in contrast, reveals low-level morphological patterns that are indicative of obfuscation strategies, such as homoglyph substitutions, repeated character segments, or anomalous symbol combinations. Together, they provide a dual-granularity structural prior that can be leveraged during URL-specific subgraph extraction and downstream representation learning. In the following section, we describe how these global graphs are utilized to initialize and encode structural context for individual URL samples.

\subsection{URL Sequence Encoder and Graph Encoder}

To effectively capture both semantic-level and morphological-level cues from a given URL, we employ a hybrid dual-branch architecture consisting of a \textit{URL sequence encoder} and a \textit{graph encoder}. During training, each input URL is encoded via two parallel branches. The first branch encodes the URL sequence using BERT and CharCNN to obtain global semantic and morphological representations. In the second branch, we extract URL-specific subgraphs from the global subword-level and character-level co-occurrence graphs described previously. The nodes in these subgraphs are initialized with contextualized features derived from the CharCNN encoding of the input URL, and further refined through a GCN-based structural encoder. This dual-pathway encoding enables our model to integrate fine-grained sequential and structural signals for enhanced URL representation learning.

\subsubsection{URL Sequence Encoder}

Given a raw input URL $u$, we apply both the subword tokenizer $\tau_t(u)$ and the character tokenizer $\tau_c(u)$ to obtain two token sequences:
\begin{equation}
\begin{aligned}
S_t &= \tau_t(u) = [w_1, w_2, \dots, w_{L_t}], \\
S_c &= \tau_c(u) = [c_1, c_2, \dots, c_{L_c}],
\end{aligned}
\end{equation}

where $L_t$ and $L_c$ are the lengths of the subword and character sequences, respectively.

We first encode the subword sequence $S_t$ using a pre-trained BERT model, which employs multi-head self-attention to produce context-aware representations for each subword token. Specifically, for each attention head, the attention weights are computed as:
\begin{equation}
\text{Attention}(Q, K, V) = \text{softmax} \left( \frac{QK^\top}{\sqrt{d}} \right) V,
\end{equation}
where $Q$, $K$, and $V$ are the query, key, and value matrices derived from the input embeddings, and $d$ is the dimension of each attention head. The outputs from all attention heads are concatenated and passed through a feedforward projection layer to yield the final token embeddings:
\begin{equation}
H_t = \text{BERT}(S_t) \in \mathbb{R}^{L_t \times d_t}.
\end{equation}

To complement the semantic-level representation from BERT, we utilize a character-level convolutional neural network (CharCNN) to encode $S_c$ and capture local morphological patterns. Let $E_c \in \mathbb{R}^{L_c \times d_c}$ denote the initial character embedding matrix. A series of 1D convolutional filters $\{W^{(k)} \in \mathbb{R}^{k \times d_c}\}$ of varying kernel widths $k$ are applied across $E_c$ to extract $n_k$-dimensional local features:
\begin{equation}
\begin{gathered}
f_i^{(k)} = \text{ReLU}(W^{(k)} * E_c[i:i+k-1] + b^{(k)}), \quad \\ i = 1, \dots, L_c - k + 1,
\end{gathered}
\end{equation}
where $*$ denotes the 1D convolution operation. A max-over-time pooling is applied to obtain the most salient feature for each kernel size:
\begin{equation}
z^{(k)} = \max_{i} f_i^{(k)}.
\end{equation}
The final CharCNN embedding of the URL is formed by concatenating all pooled features:
\begin{equation}
h_c = [z^{(k_1)} \, \| \, z^{(k_2)} \, \| \dots \| \, z^{(k_K)}] \in \mathbb{R}^{d_c'},
\end{equation}
where $d_c'$ is the sum of output dimensions from all kernels.

\subsubsection{Graph Encoder}

Following sequence encoding, we extract two URL-specific subgraphs—one from the global subword graph $\mathcal{G}_\text{word}$ and one from the character graph $\mathcal{G}_\text{char}$—based on the tokens present in $S_t$ and $S_c$, respectively. These induced subgraphs are denoted as:
\begin{equation}
\mathcal{G}_u^\text{word} = (\mathcal{V}_u^\text{word}, \mathcal{E}_u^\text{word}), \quad
\mathcal{G}_u^\text{char} = (\mathcal{V}_u^\text{char}, \mathcal{E}_u^\text{char}),
\end{equation}
where $\mathcal{V}_u^\text{word} \subseteq \mathcal{V}_t$ and $\mathcal{V}_u^\text{char} \subseteq \mathcal{V}_c$ are the token and character nodes that appear in the given URL $u$.

To initialize node features in both subgraphs, we extract the corresponding character-level subsequences for each node $v_i$ appearing in the input URL and encode them using the same CharCNN module described previously. For each node $v_i$, let $\text{seq}(v_i)$ denote the character sequence associated with $v_i$ in the input. Then the initial node embedding is defined as:
\begin{equation}
h_i^{(0)} = \text{CharCNN}(\text{seq}(v_i)) \in \mathbb{R}^{d_c'},
\end{equation}
where $d_c'$ is the same as defined in the previous section.

To encode the structural dependencies among nodes, we apply Graph Convolutional Networks (GCNs) over both subgraphs. For each layer $\ell$, the node embeddings are updated as:
\begin{equation}
h_i^{(\ell+1)} = \sigma \left( \sum_{j \in \mathcal{N}(i)} \frac{1}{\sqrt{d_i d_j}} W^{(\ell)} h_j^{(\ell)} \right),
\end{equation}
where $\mathcal{N}(i)$ denotes the neighbors of node $i$, $d_i$ is the degree of node $i$, $W^{(\ell)}$ is the trainable weight matrix at layer $\ell$, and $\sigma(\cdot)$ is a non-linear activation function. The final node representations are aggregated using average pooling to produce a graph-level representation:

\begin{equation}
\begin{aligned}
h^\text{word} &= \frac{1}{|\mathcal{V}_u^\text{word}|} \sum_{i \in \mathcal{V}_u^\text{word}} h_i^{(L)} , \quad \\
h^\text{char} &= \frac{1}{|\mathcal{V}_u^\text{char}|} \sum_{i \in \mathcal{V}_u^\text{char}} h_i^{(L)} ,
\end{aligned}
\end{equation}

 These two graph-level embeddings provide complementary structural characterizations of the input URL at the subword and character granularity levels, respectively. In the subsequent section, we describe how these representations are integrated with the sequence-level features to form a unified encoding for downstream classification.

\subsection{Dynamic Multi-modal Fusion and Joint Optimization}

To comprehensively capture both the semantic and structural characteristics of a URL $u$, we dynamically fuse three representations: (1) the sequence-level embedding $h_t$ from the BERT encoder, (2) the subword-level graph embedding $h_u^{\text{word}}$ from the subgraph $\mathcal{G}_u^{\text{word}}$, and (3) the character-level graph embedding $h_u^{\text{char}}$ from the subgraph $\mathcal{G}_u^{\text{char}}$. The fusion is performed in a gated manner, allowing the model to learn instance-specific weighting between the sequential and structural signals.

\subsubsection*{Feature Alignment and Gated Fusion Mechanism}

Since the graph-based representations $h_u^{\text{word}} \in \mathbb{R}^{d_g}$ and $h_u^{\text{char}} \in \mathbb{R}^{d_g}$ differ in dimension from the BERT output $h_t \in \mathbb{R}^{d_t}$, we apply two linear projection layers to map them into the same latent space of dimension $d_t$:
\begin{equation}
\tilde{h}_u^{\text{word}} = W_{\text{word}} h_u^{\text{word}} + b_{\text{word}}, \quad 
\tilde{h}_u^{\text{char}} = W_{\text{char}} h_u^{\text{char}} + b_{\text{char}},
\end{equation}
where $W_{\text{word}}, W_{\text{char}} \in \mathbb{R}^{d_t \times d_g}$ and $b_{\text{word}}, b_{\text{char}} \in \mathbb{R}^{d_t}$ are learnable parameters.

The projected vectors are concatenated with the original BERT embedding to form a unified feature vector:
\begin{equation}
h_u^{\text{concat}} = [h_t \ \| \ \tilde{h}_u^{\text{word}} \ \| \ \tilde{h}_u^{\text{char}}] \in \mathbb{R}^{3d_t}.
\end{equation}

A gating network is applied to dynamically compute the relative importance of the three representations. Specifically, a soft attention weight vector is generated as:
\begin{equation}
\boldsymbol{\alpha} = \text{softmax}(W_g h_u^{\text{concat}} + b_g) \in \mathbb{R}^{3},
\end{equation}
where $W_g \in \mathbb{R}^{3 \times 3d_t}$ and $b_g \in \mathbb{R}^{3}$ are learnable parameters. The final fused representation is then computed as a convex combination:
\begin{equation}
h_u^{\text{fused}} = \alpha_1 h_t + \alpha_2 \tilde{h}_u^{\text{sub}} + \alpha_3 \tilde{h}_u^{\text{char}} \in \mathbb{R}^{d_t},
\end{equation}
where $\alpha_1 + \alpha_2 + \alpha_3 = 1$ and $\alpha_i \geq 0$ for $i=1,2,3$.

\subsubsection*{Joint Training and Efficient Batched Optimization}

The fused representation $h_u^{\text{fused}}$ is passed through a classification layer:
\begin{equation}
\hat{y}_u = \text{softmax}(W_c h_u^{\text{fused}} + b_c), \quad W_c \in \mathbb{R}^{C \times d_t}, \quad b_c \in \mathbb{R}^{C},
\end{equation}
where $C$ denotes the number of output classes.

Model training is supervised by minimizing the cross-entropy loss:
\begin{equation}
\mathcal{L}_{\text{u}} = -\sum_{u} \sum_{c=1}^{C} \mathbf{1}_{[y_u = c]} \log \hat{y}_{u,c}.
\end{equation}

The parameters jointly optimized during training include those of the character-level CNN encoder, the two graph convolutional networks (for $\mathcal{G}_u^{\text{word}}$ and $\mathcal{G}_u^{\text{char}}$), the projection layers $W_{\text{sub}}, W_{\text{char}}$, the gating network parameters $W_g, b_g$, and the final classification layer. The BERT encoder is optionally fine-tuned depending on training configuration.

To ensure training efficiency and scalability, we adopt batched graph processing via a block-based mini-batching strategy. For each training batch $\mathcal{B} = \{u_1, u_2, \dots, u_B\}$ consisting of $B$ URLs, we construct the corresponding subgraphs $\{\mathcal{G}_{u_i}^{\text{word}}\}_{i=1}^B$ and $\{\mathcal{G}_{u_i}^{\text{char}}\}_{i=1}^B$ for the subword- and character-level views, respectively. These subgraphs are merged into two batched graphs:
\begin{equation}
\begin{aligned}
\mathcal{G}^{\text{word}}_{\text{batch}} &= \text{batch}(\{\mathcal{G}_{u_i}^{\text{word}}\}_{i=1}^B), \quad \\
\mathcal{G}^{\text{char}}_{\text{batch}} &= \text{batch}(\{\mathcal{G}_{u_i}^{\text{char}}\}_{i=1}^B),
\end{aligned}
\end{equation}
where $\text{batch}(\cdot)$ denotes the disjoint union operation with batched graph metadata preserved for downstream pooling.

Each node in the batched graph is associated with initial features and GCN layers are applied in parallel across the batch. After message passing, node-level outputs are aggregated using mean pooling per subgraph:
\begin{equation}
\begin{aligned}
&h_{u_i}^{\text{word}} = \text{mean-pool}\left(\text{GCN}_{\text{word}}(\mathcal{G}^{\text{word}}_{\text{batch}})\right)_{u_i}, \quad \\
&h_{u_i}^{\text{char}} = \text{mean-pool}\left(\text{GCN}_{\text{char}}(\mathcal{G}^{\text{char}}_{\text{batch}})\right)_{u_i}.
\end{aligned}
\end{equation}
These graph embeddings are fused with the corresponding BERT representation $h_{t_i}$ as previously described to produce the fused representation $h_{u_i}^{\text{fused}}$ for each URL $u_i$ in the batch.

The batch-level loss is computed as the average cross-entropy over all samples:
\begin{equation}
\mathcal{L}_{\text{batch}} = \frac{1}{B} \sum_{i=1}^{B} \mathcal{L}_{u_i} = 
- \frac{1}{B} \sum_{i=1}^{B} \sum_{c=1}^{C} \mathbf{1}_{[y_{u_i} = c]} \log \hat{y}_{u_i,c}.
\end{equation}

This unified training strategy enables joint end-to-end optimization of all model components—including the BERT encoder, graph-based encoders, and gating network—using backpropagation. The integration of global semantic, morphological, and structural features allows the model to generalize effectively, even under limited supervision, and robustly encode the heterogeneous properties of URLs.

\section{Datasets}
\begin{table*}[ht]
\small
\centering
\caption{Overall and class-specific distribution of Top-Level Domains (TLDs).}
\label{tab:dataset_tlds}
\setlength{\tabcolsep}{7.5pt}
\renewcommand{\arraystretch}{1.20}
\begin{tabular}{l
                lll
                lll
                lll}
\toprule
\multirow{2}{*}{\textbf{Dataset}} &
\multicolumn{3}{c}{\textbf{Total TLDs }} &
\multicolumn{3}{c}{\textbf{Benign TLDs }} &
\multicolumn{3}{c}{\textbf{Malicious TLDs }} \\
\cmidrule(lr){2-4}\cmidrule(lr){5-7}\cmidrule(lr){8-10}
& .com & ccTLDs & other gTLDs
& .com & ccTLDs & other gTLDs
& .com & ccTLDs & other gTLDs \\
\midrule
GramBeddings\footnotemark[1] & 56.14\% & 11.93\% & 31.93\% & 52.17\% & 12.04\% & 35.79\% & 60.10\% & 11.82\% & 28.08\% \\
Mendeley\footnotemark[2]     & 62.22\% & 0.95\%  & 36.83\% & 61.97\% & 0.93\%  & 37.10\% & 72.86\% & 1.61\%  & 25.53\% \\
Kaggle\footnotemark[3]      & 64.03\% & 5.62\%  & 30.35\% & 77.46\% & 0.63\%  & 21.92\% & 50.59\% & 10.61\% & 38.80\% \\
\bottomrule
\end{tabular}
\end{table*}

\begin{table}[ht]
\small
\centering
\caption{Sample statistics of the datasets used in our experiments. }
\label{tab:dataset_counts}
\setlength{\tabcolsep}{10pt}
\renewcommand{\arraystretch}{1.15}
\begin{tabular}{lccc}
\toprule
\textbf{Dataset} & \textbf{Total} & \textbf{Benign} & \textbf{Malicious} \\
\midrule
GramBeddings & 800{,}000 & 400{,}000 & 400{,}000 \\
Mendeley     & 1{,}561{,}934 & 1{,}526{,}619 & 35{,}315 \\
Kaggle      & 632{,}503 & 316{,}252 & 316{,}251 \\
\bottomrule
\end{tabular}
\end{table}

To comprehensively evaluate the proposed \textit{URL2Graph++} framework, we utilize three publicly available datasets that differ significantly in terms of sample size, class balance, and Top-Level Domain (TLD) distributions. These datasets are selected to test the model under complementary conditions, including class-balanced settings, extreme class imbalance, and cross-dataset validation. Detailed dataset statistics are summarized in Table~\ref{tab:dataset_tlds} and Table~\ref{tab:dataset_counts}.

\textbf{GramBeddings Dataset.}  
The GramBeddings dataset~\cite{bozkir2023grambeddings} contains 800,000 samples evenly divided between 400,000 malicious and 400,000 benign URLs. Malicious URLs were collected from repositories such as PhishTank and OpenPhish, while benign URLs were obtained by crawling Alexa-ranked websites with down-sampling applied to ensure balance. Within benign URLs, 52.17\% use the .com TLD, 12.04\% are ccTLDs, and 35.79\% are other gTLDs. Malicious URLs are distributed as 60.10\% .com, 11.82\% ccTLDs, and 28.08\% other gTLDs. This dataset serves as a controlled class-balanced benchmark, allowing us to evaluate the model's detection performance in scenarios where benign and malicious samples are evenly represented.

\textbf{Mendeley Dataset.}  
The Mendeley dataset~\cite{singh2020malicious} comprises 1,561,934 samples, including 1,526,619 benign URLs and 35,315 malicious URLs, yielding a highly imbalanced ratio of approximately 1:43. The URLs were collected with the MalCrawler tool and validated using the Google Safe Browsing API. Benign URLs predominantly use the .com TLD (61.97\%), with 0.93\% in ccTLDs and 37.10\% in other gTLDs. Malicious URLs are distributed as 72.86\% .com, 1.61\% ccTLDs, and 25.53\% other gTLDs. This dataset is used to test model robustness under extreme class imbalance, reflecting real-world conditions where malicious URLs are scarce relative to benign ones.

\textbf{Kaggle Dataset.}  
The Kaggle dataset includes 632,503 samples, equally divided between 316,251 malicious and 316,252 benign URLs. Benign samples consist of 77.46\% .com, 0.63\% ccTLDs, and 21.92\% other gTLDs, while malicious samples contain 50.59\% .com, 10.61\% ccTLDs, and 38.80\% other gTLDs. Due to its balanced structure and distinct distribution patterns compared to GramBeddings and Mendeley, this dataset is employed to serve as a cross-dataset validation benchmark, enabling us to assess the generalization ability of our model across different sources.

The combination of these datasets provides a diverse and rigorous evaluation setting. GramBeddings tests the model under balanced conditions, Mendeley examines robustness against severe imbalance, and Kaggle facilitates cross-dataset validation, collectively ensuring a thorough assessment of both effectiveness and generalization in malicious URL detection, providing a comprehensive foundation for evaluating URL2Graph++.

\section{Experiments}

\begin{table*}[ht]
\centering
\small
\caption{Performance comparison of our method against others using the Grambedding dataset and Mendeley dataset.}
\label{tab:results}
\setlength{\tabcolsep}{5.2pt}
\renewcommand{\arraystretch}{1.1}
\begin{tabular}{@{}c l *{11}{c}@{}}
\toprule
\multicolumn{2}{c}{\textbf{Datasets}}
 & \multicolumn{5}{c}{\textbf{Grambedding}} 
 & \multicolumn{5}{c}{\textbf{Mendeley}}
 \\
\cmidrule(lr){3-7} \cmidrule(lr){8-12} 
Training Size & Method & \textbf{Prec} & \textbf{ACC} & \textbf{Rec} & \textbf{F1} & \textbf{AUC} 
 & \textbf{Prec} & \textbf{ACC} & \textbf{Rec} & \textbf{F1} & \textbf{AUC} 
 \\
 
\midrule
\multirow{6}{*}{\makecell{1\%\\6400 (Grambedding)\\15620 (Mendeley)}}
& Dephides 
& 0.6258 & 0.6316 & 0.5847 &  0.6045 &  0.6527
& 0.7962 & 0.5013 & 0.5271 &  0.5138 &  0.6100\\
& CGAN 
& 0.6509 & 0.6195 & 0.7255 &  0.6861 &  0.7484 
& 0.9299 & 0.6337 & 0.8135 &  0.7124 &  0.7037\\
& URLBERT 
& 0.8836 & 0.9017 & 0.9124 &  0.8978 &  0.9512 
& 0.9716 & 0.5141 & 0.4829 &  0.4980 &  0.8371\\
& LLaMA-2 
& 0.9038 & 0.9237 & 0.9204 &  0.9120 &  0.9687 
& 0.9792 & 0.6249 & 0.5300 &  0.5736 &  0.8144\\
& \textbf{Our} 
& \textbf{0.9307} & \textbf{0.9318} & \textbf{0.9316}& \textbf{0.9311}& \textbf{0.9802} 
& \textbf{0.9818} & \textbf{0.8703} & \textbf{0.6712}& \textbf{0.7578}& \textbf{0.8684}\\

\midrule
\multirow{6}{*}{\makecell{2\%\\12800 (Grambedding)\\31240 (Mendeley)}}
& Dephides 
& 0.6743 & 0.6706 & 0.6602 &  0.6667 &  0.6634
& 0.9049 & 0.5132 & 0.3727 &  0.4318 &  0.7265\\
& CGAN 
& 0.7395 & 0.6811 & 0.7198 &  0.7295 &  0.8161
& 0.9336 & 0.7096 & 0.6695 &  0.8140 &  0.7622\\
& URLBERT 
& 0.8902 & 0.9047 & 0.9231 &  0.9034 &  0.9580 
& 0.9729 & 0.5097 & 0.3091 &  0.3848 &  0.8006\\
& LLaMA-2 
& 0.9141 & 0.9289 & 0.9270 &  0.9205 &  0.9702 
& 0.9594 & 0.6101 & 0.4991 &  0.5490 &  0.7902\\
& \textbf{Our} 
& \textbf{0.9378} & \textbf{0.9381} & \textbf{0.9372}& \textbf{0.9374}& \textbf{0.9867} 
& \textbf{0.9840} & \textbf{0.8482} & \textbf{0.7361}& \textbf{0.7881}& \textbf{0.8712}\\
\midrule

\multirow{6}{*}{\makecell{3\%\\19200 (Grambedding)\\46860 (Mendeley)}}
& Dephides 
& 0.6952 & 0.7180 & 0.6801 &  0.6880 &  0.7237
& 0.9401 & 0.5398 & 0.4662 &  0.5003 &  0.7332\\
& CGAN 
& 0.7752 & 0.7442 & 0.8128 &  0.7935 &  0.8721
& 0.7885 & 0.5117 & 0.6206 &  0.5609 &  0.7381\\
& URLBERT 
& 0.8997 & 0.9100 & 0.9267 &  0.9130 &  0.9644 
& 0.9724 & 0.5201 & 0.4762 &  0.4971 &  0.8457\\
& LLaMA-2 
& 0.9192 & 0.9339 & 0.9326 &  0.9259 &  0.9741
& 0.9645 & 0.6302 & 0.5231 &  0.5716 &  0.8262\\
& \textbf{Our} 
& \textbf{0.9443} & \textbf{0.9444} & \textbf{0.9402}& \textbf{0.9422}& \textbf{0.9903} 
& \textbf{0.9856} & \textbf{0.9287} & \textbf{0.7027}& \textbf{0.8204}& \textbf{0.8744}\\
\midrule

\multirow{6}{*}{\makecell{4\%\\25600 (Grambedding)\\62840 (Mendeley)}}
& Dephides 
& 0.7031 & 0.7372 & 0.6883 &  0.6959 &  0.7456
& 0.8931 & 0.6211 & 0.5072 &  0.5584 &  0.7237\\
& CGAN 
& 0.8395 & 0.7927 & 0.8847 &  0.8614 &  0.8970 
& 0.8736 & 0.4022 & 0.5048 &  0.4476 &  0.6767\\
& URLBERT 
& 0.9126 & 0.9188 & 0.9300 &  0.9212 &  0.9699 
& 0.9791 & 0.5441 & 0.5002 &  0.5212 &  0.8671\\
& LLaMA-2 
& 0.9245 & 0.9401 & 0.9380 &  0.9312 &  0.9800
& 0.9711 & 0.6053 & 0.5921 &  0.5986 &  0.8554\\
& \textbf{Our} 
& \textbf{0.9467} & \textbf{0.9475} & \textbf{0.9463}& \textbf{0.9467}& \textbf{0.9921}
& \textbf{0.9794} & \textbf{0.7653} & \textbf{0.7607}& \textbf{0.7629}& \textbf{0.8837}\\
\midrule

\multirow{6}{*}{\makecell{5\%\\32000 (Grambedding)\\78100 (Mendeley)}}
& Dephides 
& 0.7107 & 0.7416 & 0.6917 &  0.7011 &  0.7987
& 0.9320 & 0.6351 & 0.6771 &  0.6554 &  0.7633\\
& CGAN 
& 0.8830 & 0.8461 & 0.8962 &  0.8898 &  0.9047
& 0.8418 & 0.7491 & 0.7928 &  0.7703 &  0.8895\\
& URLBERT 
& 0.9194 & 0.9179 & 0.9345 &  0.9269 &  0.9719
& 0.9769 & 0.5279 & 0.4982 &  0.5126 &  0.8444\\
& LLaMA-2 
& 0.9298 & 0.9457 & 0.9377 &  0.9337 &  0.9824
& 0.9745 & 0.6231 & 0.6001 &  0.6114 &  0.8692\\
& \textbf{Our} 
& \textbf{0.9482} & \textbf{0.9490} & \textbf{0.9486}& \textbf{0.9484}& \textbf{0.9943}
& \textbf{0.9851} & \textbf{0.8899} & \textbf{0.7302}& \textbf{0.8022}& \textbf{0.8903}\\

\bottomrule
\end{tabular}
\end{table*}

In this section, we delineate a comprehensive and structured experimental protocol to assess the efficacy of our proposed approach.  We design and conduct five types of experiments. Each is intended to assess a specific aspect of model behavior across different dataset distributions, attack conditions, and input variations.

\begin{itemize}
    \item \textbf{Benchmark Evaluation.} We evaluate the classification performance of our model against a set of competitive baselines on two datasets: one with relatively balanced label distribution and another with significant class imbalance. Metrics such as accuracy, precision, recall, and F1-score are used to assess overall effectiveness.
    
    \item \textbf{Cross-Dataset Generalization.} To examine the model’s transferability and generalization capacity, we train on one dataset and evaluate on another with different statistical properties and label distribution, without fine-tuning on the target domain.
    
    \item \textbf{Ablation Study.} We systematically ablate key architectural components—including the graph encoder, CharCNN modules, and the Dynamic Multi-modal Fusion mechanism—to quantify their individual contribution to overall performance.
    
    \item \textbf{Adversarial Robustness.} We assess the resilience of the model under adversarial perturbations by evaluating its detection accuracy against synthetically generated adversarial URLs, focusing on true positive rate (TPR) across varying false positive rate (FPR) thresholds.
    
    \item \textbf{Short URL Detection.} To further evaluate generalization to sparse lexical patterns, we test the model on a subset of short URLs, characterized by reduced token length, and compare performance degradation relative to longer samples.
\end{itemize}

\noindent
The above experimental configurations allow us to evaluate the model from multiple perspectives, including in-distribution accuracy, cross-domain adaptability, component-wise effectiveness, resistance to input perturbations, and robustness to limited lexical content.

\medskip
\footnotetext[1]{\url{https://web.cs.hacettepe.edu.tr/~selman/grambeddings-dataset/}}
\footnotetext[2]{\url{https://data.mendeley.com/datasets/gdx3pkwp47/2}}
\footnotetext[3]{\url{https://www.kaggle.com/datasets/samahsadiq/benign-and-malicious-urls}}
\noindent\textbf{Baselines.} To benchmark the proposed model, we compare it with several state-of-the-art malicious URL detection methods, including both sequence-based and hybrid architectures. The selected baselines include Dephides\cite{sahingoz2024dephides}, CGAN\cite{kamran2021cgan}, URLBERT\cite{li2025urlbert} and LLaMA-2\cite{touvron2023llama}  that have demonstrated strong performance on URL classification benchmarks. These models cover a range of architectural designs, from purely sequential encoders to those incorporating external features or graph-enhanced representations. All baseline implementations follow the settings reported in their respective publications, and are trained and evaluated under the same experimental conditions as our model for fair comparison.
\begin{table}[t]
\centering
\small
\caption{TPR@FPR Evaluation of baselines and our model on Grambedding dataset (Training Size = 6400).}
\label{tab:special_6400}
\renewcommand{\arraystretch}{1.1}
\setlength{\tabcolsep}{6.5pt}
\begin{tabular}{@{}lccccc@{}}
\toprule
& \multicolumn{4}{c}{\textbf{TPR @ FPR Level}} & \\
\cmidrule(lr){2-5}
\textbf{Method}  & 0.0001 & 0.001 & 0.01 & 0.1 & \textbf{AUC}  \\
\midrule
Dephides       & 0.0000 & 0.0038 & 0.0825 & 0.5472 & 0.6527 \\
CGAN & 0.0127 & 0.0866 & 0.4371 & 0.7092 & 0.7487 \\
URLBERT        & 0.1453 & 0.3900 & 0.6016 & 0.9121 & 0.9512 \\
LLaMA-2        & 0.1190 & 0.3491 & 0.5874 & 0.9124 & 0.9687 \\
\midrule
\textbf{Our}   & \textbf{0.1739} & \textbf{0.4073} & \textbf{0.6324} & \textbf{0.9431} & \textbf{0.9802} \\
\bottomrule
\end{tabular}
\end{table}

\subsection{Comparison Experiment}

\paragraph{Grambedding dataset}
To examine the scalability of the proposed method, we first evaluate binary classification performance on the balanced Grambedding dataset with different training sizes. As shown in Table \ref{tab:results}, across all proportions of the training set, a consistent trend emerges: as the amount of training data increases, the performance of all methods improves in terms of accuracy, precision, recall, F1-score, and AUC. However, the proposed method maintains a clear advantage over the baselines at every scale. 

At the smallest training set size of 6,400 samples, the proposed method achieves an accuracy of 0.9318 and an AUC of 0.9802, clearly outperforming Dephides (accuracy 0.6316) and CGAN (accuracy 0.6195). With increased training size, the gains become more evident. For instance, at 32,000 samples, the proposed method yields an accuracy of 0.9490 and an AUC of 0.9943, which not only exceeds Dephides and CGAN by a wide margin but also surpasses stronger baselines such as URLBERT and LLaMA-2. Although URLBERT and LLaMA-2 show stable improvements across scales, their peak accuracies (0.9179 and 0.9457 respectively at 32,000 samples) remain slightly inferior. These results demonstrate that the proposed method achieves both higher efficiency with limited data and stronger scalability when provided with larger datasets.

\paragraph{Mendeley dataset.}
We further assess performance on the highly imbalanced Mendeley dataset, which better reflects real-world distributions where benign URLs vastly outnumber malicious ones. As shown in Table \ref{tab:results}, all methods are negatively affected by the severe class imbalance, but the proposed method exhibits greater robustness across training sizes.

At 1\% of the dataset, the proposed method achieves an F1-score of 0.7578 and an AUC of 0.8684, substantially higher than Dephides (F1-score 0.5138) and CGAN (F1-score 0.7124). As the training set grows, the advantage becomes increasingly evident. For example, at 78,100 samples, the proposed method attains an F1-score of 0.8022 and an AUC of 0.8903, whereas URLBERT and LLaMA-2 struggle to reach comparable performance due to higher false positive rates. The consistent advantage of the proposed method across varying training scales highlights its ability to capture discriminative features even under severe imbalance, thereby maintaining a favorable trade-off between precision and recall.

\paragraph{TPR@FPR Evaluation.}
To further examine robustness, we compare models at different false positive rate (FPR) thresholds, focusing on their true positive rate (TPR) and AUC values. Table~\ref{tab:special_6400} and \ref{tab:special_32000} show that the proposed method achieves consistently superior TPR across all FPR levels and training sizes on Grambedding dataset.

\begin{table}[t]
\centering
\small
\caption{TPR@FPR Evaluation of baselines and our model on Grambedding dataset (Training Size = 32000).}
\label{tab:special_32000}
\renewcommand{\arraystretch}{1.1}
\setlength{\tabcolsep}{6.5pt}
\begin{tabular}{@{}lccccc@{}}
\toprule
 & \multicolumn{4}{c}{\textbf{TPR @ FPR Level}} &  \\
\cmidrule(lr){2-5}
\textbf{Method}  & 0.0001 & 0.001 & 0.01 & 0.1 & \textbf{AUC} \\
\midrule
Dephides       & 0.0024 & 0.0316 & 0.2162 & 0.6746 & 0.7987 \\
CGAN & 0.1350 & 0.7049 & 0.8302 & 0.8824 & 0.9047 \\
URLBERT        & 0.1671 & 0.4697 & 0.7962 & 0.9021 & 0.9719 \\
LLaMA-2        & 0.1331 & 0.5073 & 0.7812 & 0.9395 & 0.9824 \\
\midrule
\textbf{Our}   & \textbf{0.1466} & \textbf{0.5034} & \textbf{0.8001} & \textbf{0.9737} & \textbf{0.9943} \\
\bottomrule
\end{tabular}
\end{table}

At a training size of 6,400, the proposed method achieves a TPR of 0.1739 at an FPR of 0.0001, whereas baselines such as Dephides and CGAN fail to maintain meaningful detection capability under the same condition. As the training size increases to 32,000, the proposed method continues to demonstrate clear advantages, achieving a TPR of 0.5034 at FPR 0.001 and maintaining an AUC of 0.9943, surpassing the strongest baseline LLaMA-2 (AUC 0.9824). These results suggest that the proposed method is more effective at identifying malicious URLs under strict low-FPR requirements, an essential property in security-sensitive applications where false alarms must be minimized.

The comprehensive results across balanced and imbalanced datasets, as well as under stringent low-FPR evaluation, consistently indicate that the proposed method achieves clear advantages over baseline methods. On balanced data, the method attains higher accuracy and AUC compared with strong competitors, suggesting enhanced ability to learn from sufficient supervision. On highly imbalanced data, the method maintains stable precision–recall trade-offs, highlighting robustness in scenarios where benign samples dominate. Moreover, under strict low-FPR thresholds, the method sustains markedly higher TPR than competing methods, indicating effectiveness in minimizing false negatives while keeping false alarms low. Collectively, these findings demonstrate that the proposed method delivers both efficiency and robustness, supporting its practical utility in large-scale malicious URL detection.

\begin{figure}[t]
    \centering
    \includegraphics[width=0.49\textwidth]{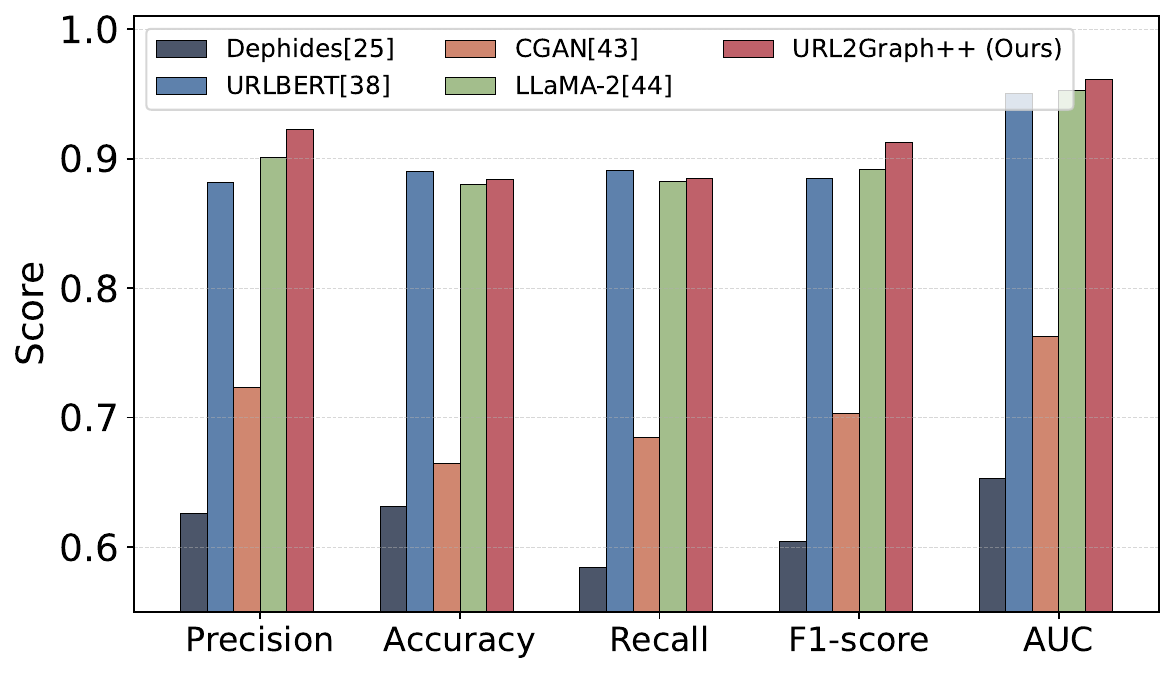}
    \caption{Cross-dataset evaluation where models are trained on the Grambedding dataset and tested on the Kaggle dataset. Results are reported on five metrics (Precision, Accuracy, Recall, F1-score, and AUC). }
    \label{fig:gram2kaggle}
\end{figure}

\begin{figure}[t]
    \centering
    \includegraphics[width=0.49\textwidth]{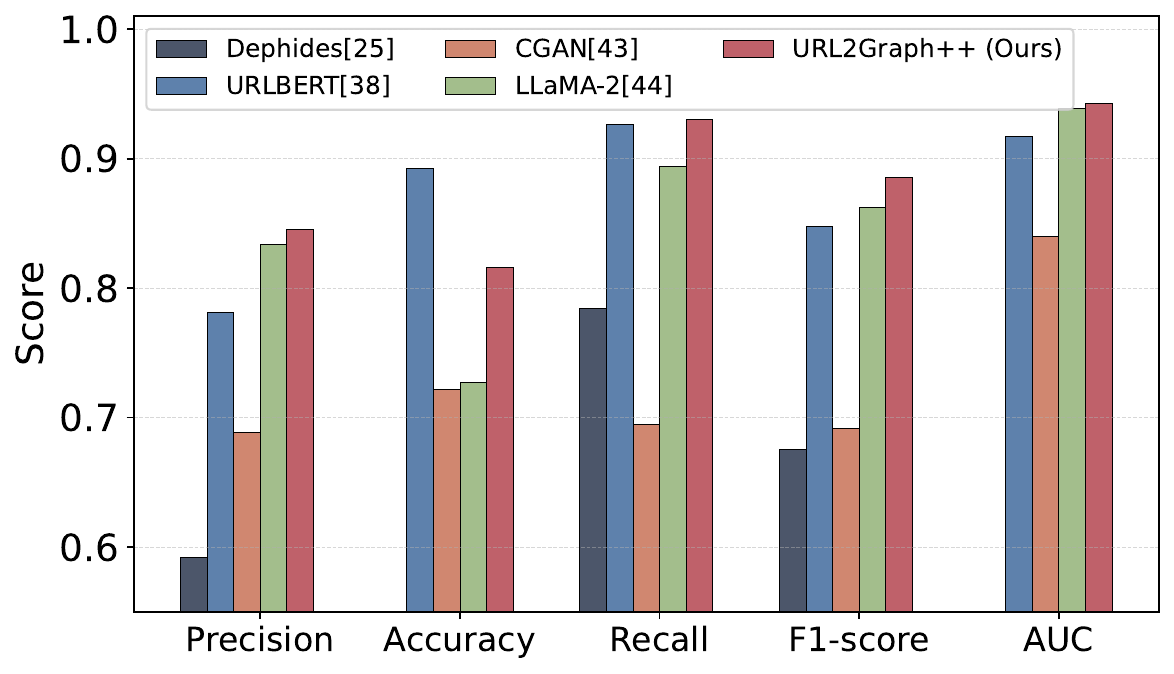}
    \caption{Cross-dataset evaluation where models are trained on the Kaggle dataset and tested on the Grambedding dataset. Results are reported on five metrics (Precision, Accuracy, Recall, F1-score, and AUC). }
    \label{fig:kaggle2gram}
\end{figure}

\subsection{Cross-Dataset Validation}

To further investigate the generalization ability of the proposed model, we conduct cross-dataset validation experiments. Specifically, we train the models on the Grambedding\_dataset and evaluate them on the Kaggle\_binary\_dataset, and vice versa. For both settings, 30\% of the dataset is used for training and 5\% for validation. The results are summarized in Figure \ref{fig:gram2kaggle} and \ref{fig:kaggle2gram}.

When trained on the Grambedding dataset and tested on the Kaggle binary dataset, our model consistently achieves the best performance across all metrics, reaching a precision of 0.9224, recall of 0.8845, F1-score of 0.9127, and an AUC of 0.9613. In comparison, URLBERT and LLaMA-2 achieve competitive results, particularly in precision and AUC, yet still fall slightly behind our method in terms of accuracy and F1-score. CGAN and Dephides, on the other hand, exhibit significantly lower recall and AUC, highlighting their weaker ability to generalize across datasets.  In the reverse setting, where models are trained on the Kaggle\_binary\_dataset and evaluated on the Grambedding\_dataset, our approach again demonstrates superior generalization. It attains the highest recall (0.9301) and F1-score (0.8857), along with an AUC of 0.9426, surpassing all baselines. While URLBERT and LLaMA-2 also maintain strong performance in this configuration, they do not match the balanced trade-off between precision, recall, and AUC achieved by our model. By contrast, CGAN and Dephides exhibit considerable drops in accuracy and AUC, confirming their limited adaptability to unseen data distributions.

Overall, these results underscore the robustness and transferability of the proposed model. Unlike baselines that show notable degradation when confronted with domain shifts, our method consistently preserves high accuracy, recall, and AUC across datasets. This indicates that the features captured by our model are not only discriminative but also resilient to variations in data sources, thereby ensuring stronger adaptability to evolving malicious URL detection scenarios in practice.
\subsection{Ablation Experiments}

\begin{table*}[ht]
\centering
\small
\caption{Ablation study on model components, showing the performance of each variant and the relative performance degradation compared with the complete model.}
\label{tab:ablation}
\setlength{\tabcolsep}{7.1pt}
\renewcommand{\arraystretch}{1.15}
    \begin{tabular}{l c c c c c}
    \toprule
    \multicolumn{1}{c}{Method} & 
    \multicolumn{1}{c}{Accuracy} & 
    \multicolumn{1}{c}{Precision} & 
    \multicolumn{1}{c}{Recall} &
    \multicolumn{1}{c}{F1\_score} &
    \multicolumn{1}{c}{AUC}\\
    \midrule
    Graph Only
    & 0.5005 (47.22\%\textdownarrow) & 0.5007 (47.24\%\textdownarrow) & 0.4982 (47.48\%\textdownarrow) & 0.4994 (47.37\%\textdownarrow) & 0.5277 (46.93\%\textdownarrow)\\
    BERT Only
    & 0.8783 (7.37\%\textdownarrow) &  0.8943 (5.76\%\textdownarrow) 
    & 0.8779 (7.45\%\textdownarrow) &  0.8860 (6.62\%\textdownarrow)
    & 0.9445 (5.01\%\textdownarrow)
    \\
    Graph+BERT 
    & 0.9282 (2.11\%\textdownarrow) &  0.8564 (9.76\%\textdownarrow)
    & 0.8254 (12.99\%\textdownarrow) &  0.8406 (11.40\%\textdownarrow) 
    & 0.8976 (9.73\%\textdownarrow)
    \\
    Graph+BERT+CharCNN 
    & 0.9178 (3.21\%\textdownarrow) &  0.9184 (3.22\%\textdownarrow)
    & 0.9177 (3.26\%\textdownarrow) &  0.9180 (3.25\%\textdownarrow) 
    & 0.9637 (3.08\%\textdownarrow)
    \\
    Graph+BERT+DynMM 
    & 0.9195 (3.03\%\textdownarrow) &  0.9201 (3.05\%\textdownarrow)
    & 0.9196 (3.06\%\textdownarrow) &  0.9198 (3.06\%\textdownarrow) 
    & 0.9832 (1.12\%\textdownarrow)
    \\ 
    \midrule
    \textbf{Ours} 
    &\textbf{0.9482} &\textbf{0.9490} &\textbf{0.9486} &\textbf{0.9488} &\textbf{0.9943}
    \\
    \bottomrule
    \end{tabular}
\end{table*}

To systematically evaluate the contribution of each module in our framework, we adopt a stepwise ablation strategy, progressively adding components to construct five controlled variants in addition to the full model. The first variant, \emph{Graph Only}, retains only the dual-granularity URL graphs, where node features are randomly initialized and concatenated for classification; this serves as a minimal structural baseline without semantic or character-level information. The second variant, \emph{BERT Only}, removes the graph pathway and uses only the BERT encoder to derive semantic representations of URL tokens. The third variant, \emph{Graph+BERT}, integrates the graph-derived structural features with BERT outputs through direct concatenation, but does not yet include character-level modeling or adaptive fusion. The fourth variant, \emph{Graph+BERT+CharCNN}, introduces character-level representations from a CharCNN, which replace random initialization as node features in the URL graphs, while fusion remains static. The fifth variant, \emph{Graph+BERT+DynMM}, retains random initialization for graph nodes but employs the proposed Dynamic Multi-modal Fusion (DynMM) mechanism, which adaptively reweights BERT and graph features during integration. Finally, the complete model (\emph{Ours}) combines both character-informed node initialization and dynamic fusion, yielding the full configuration of URL2Graph++.

The results of these ablation experiments on Grambedding dataset, presented in Table~\ref{tab:ablation}, highlight several important findings. The \emph{Graph Only} variant performs close to random chance (accuracy 0.5005, AUC 0.5277), confirming that structural information without informative initialization is inadequate for malicious URL detection. Incorporating semantics in \emph{BERT Only} leads to a substantial performance increase (accuracy 0.8783, F1 0.8860, AUC 0.9445), demonstrating the strength of contextual token embeddings. Simple concatenation in \emph{Graph+BERT} raises accuracy to 0.9282, but reduces F1 and AUC compared to \emph{BERT Only}, indicating that uncalibrated multi-modal fusion can dilute effective signals. When character-level patterns are introduced in \emph{Graph+BERT+CharCNN}, the model achieves notable gains (F1 0.9180, AUC 0.9637), showing that CharCNN provides crucial morphological cues that stabilize graph-based representations. The \emph{Graph+BERT+DynMM} variant also improves performance (F1 0.9198, AUC 0.9832), underscoring the value of adaptive gating when combining heterogeneous features. The complete model achieves the best performance across all metrics (accuracy 0.9482, F1 0.9488, AUC 0.9943), reflecting the complementary benefits of semantic encoding, character-level initialization, and dynamic fusion.

Overall, this incremental analysis confirms that each component plays a distinct and indispensable role: semantics provide a strong foundation, character-level features unlock the potential of graph representations, and dynamic fusion ensures robust integration across modalities. The consistent performance improvements demonstrate that the performance of our proposed model stems from the joint contribution of character-level encoding, semantic and structural representations, and adaptive fusion mechanisms, all of which are essential for robust and accurate URL classification.

\subsection{Adversarial evaluation}

\begin{table}[t]
\centering
\small
\caption{Evaluation of baselines and our model under adversarial attack (Training Size = 9600).}
\label{tab:adversarial-results-9600}
\renewcommand{\arraystretch}{1.1}
\begin{tabular}{@{}lccccc@{}}
\toprule
 & \multicolumn{4}{c}{\textbf{TPR @ FPR Level}} &  \\
\cmidrule(lr){2-5}
\textbf{Method} & 0.0001 & 0.001 & 0.01 & 0.1 & \textbf{AUC} \\
\midrule
Dephides       & 0.0000 & 0.0000 & 0.0921 & 0.5479 & 0.6331 \\
CGAN & 0.0096 & 0.0866 & 0.3779 & 0.6500 & 0.7127 \\
URLBERT        & 0.1124 & 0.2652 & 0.5359 & 0.8561 & 0.9602 \\
LLaMA-2        & 0.1208 & 0.3057 & 0.7235 & 0.9124 & 0.9597 \\
\midrule
\textbf{Our}   & \textbf{0.1555} & \textbf{0.3750} & \textbf{0.6822} & \textbf{0.8871} & \textbf{0.9726} \\
\bottomrule
\end{tabular}
\end{table}
\begin{figure*}[t]
  \centering
  \begin{subfigure}[t]{1\columnwidth}
    \centering
    \includegraphics[width=\linewidth]{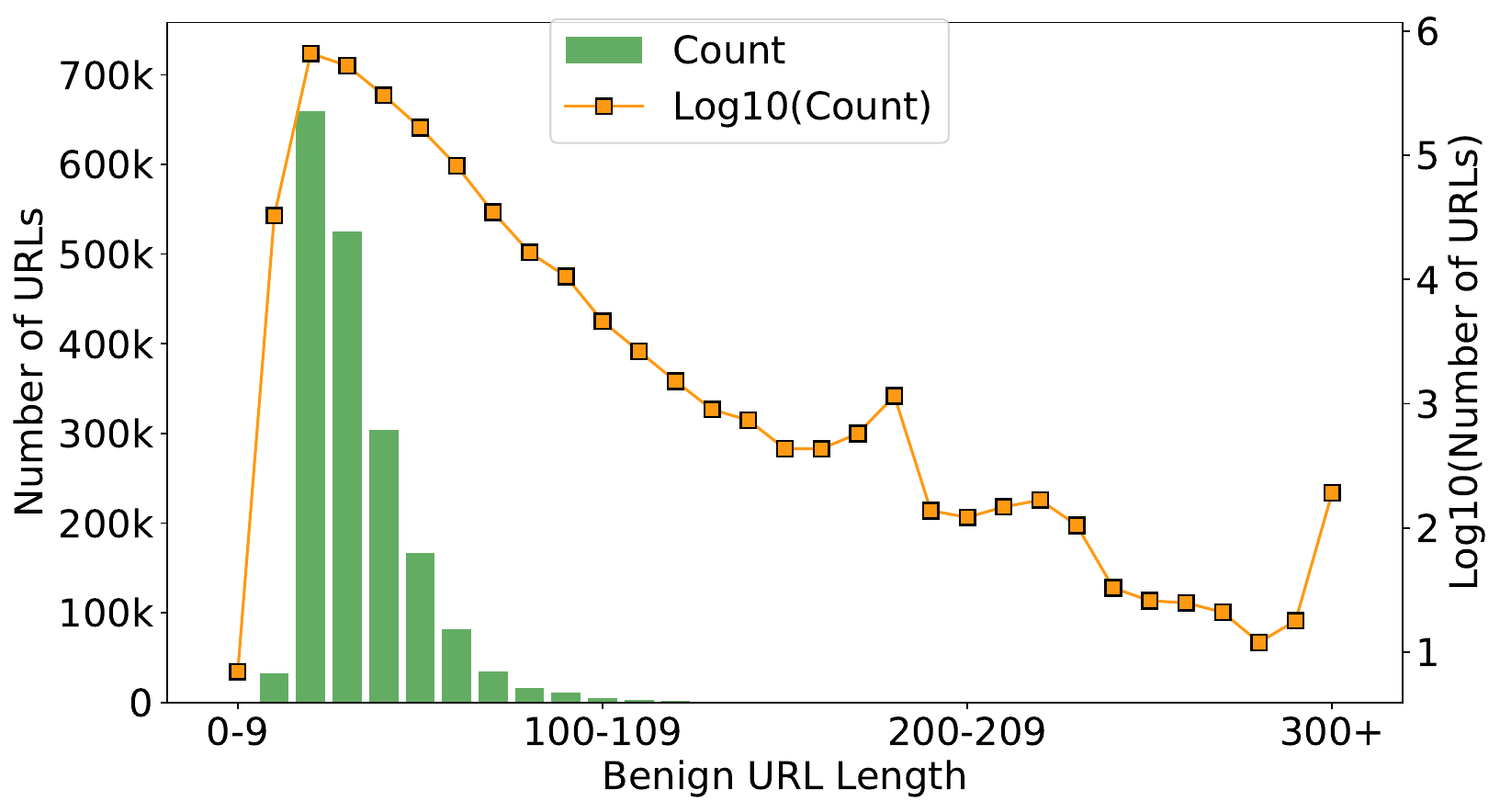}
    % \caption{Left subfigure caption.}
    \label{fig:left}
  \end{subfigure}
  \hfill
  \begin{subfigure}[t]{1\columnwidth}
    \centering
    \includegraphics[width=\linewidth]{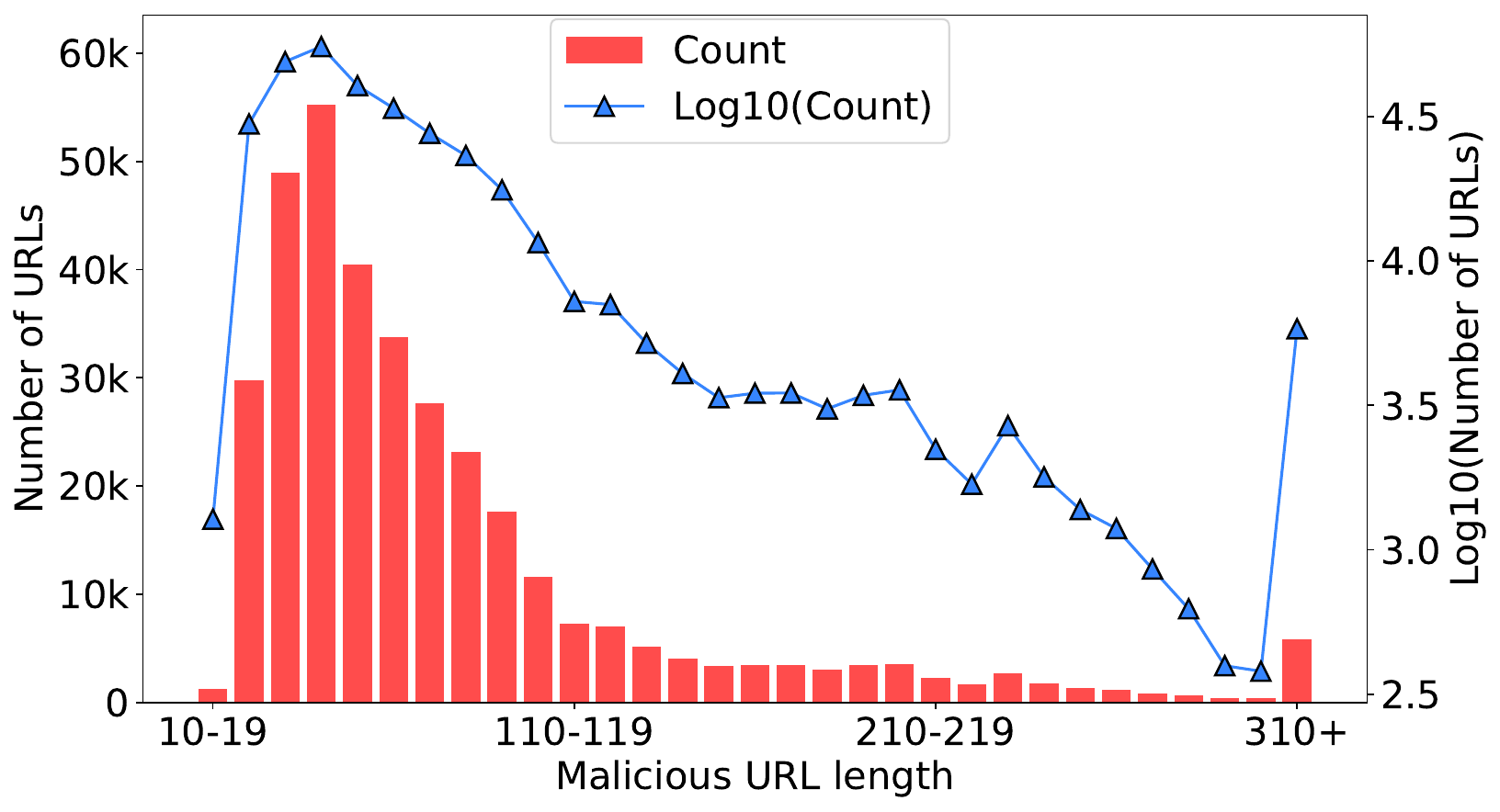}
    % \caption{Right subfigure caption.}
    \label{fig:right}
  \end{subfigure}
  \caption{Distribution of URL lengths for benign (left) and malicious (right) samples across datasets.}
  \label{fig:url_len}
\end{figure*}
Phishing attacks target short-term domains and urls that are very similar to legitimate domains and urls, which poses a significant threat. In adversarial evaluation experiments, to assess the resilience of our model to adversarial attacks, we employ a composite attack technique introduced by Grambeddings.
This approach involves creating a real-world compatible malicious URL from a legitimate example by inserting evasive characters between subword tokens within the domain name. We chose hyphens as evasive symbols. These generated domains do not inherently exist in existing training and testing datasets, but are derived from frequently observed phishing attack patterns. The process of generating adversarial samples consists of tagging the domains in a given url using XLM-RoBERTa \cite{conneau2019xlm_bert}, generating adversarial samples by tampering with the URL ratio to add letters, e.g., gooole, adding - (goo-le)
Notably, our approach involves generating a separate test dataset from the original test data. We use this new adversarial test dataset to evaluate our best performing baseline model. To visually compare our models to the baseline model, we used TPR@FPR to evaluate the results, including the value of the True positive rate for each model when the FPR is at 0.1, 0.01, 0.001, 0.0001. The experimental results summarized in Table \ref{tab:adversarial-results-9600} and \ref{tab:adversarial-results-40000} 
By comparing the results of Our model with those of the baseline model, we can see that Our model significantly outperforms the baseline model in all test situations. At a training sample of 9600, Our model's TPR (True Positive Rate) exceeds that of the baseline model at all false positive rate (FPR) levels, and especially at lower FPRs (e.g., 0.0001 and 0.001), Our model's TPR is substantially ahead. For example, at an FPR of 0.0001, the Our model has a TPR of 0.1555, while baseline models such as Dephides have a TPR close to 0, and other baseline models have significantly lower TPRs. This indicates that Our model has a significant advantage in its ability to recognise adversarial samples.The performance of Our model further improves when the training sample size is increased to 40,000. At an FPR of 0.0001, the TPR of the Our model reaches 0.1679, still outperforming all baseline models. In particular, at higher FPR levels (e.g., 0.1), the TPR of Our model is 0.9563, which is significantly higher than that of the baseline models.In terms of AUC (area under the curve), Our model has an AUC of 0.9726 with 9600 training samples, which shows stronger overall performance compared to the baseline model, e.g. 0.6331 for Dephides. Overall, Our model significantly outperforms all baseline models in terms of detection ability against samples, accuracy, and overall performance on both smaller and large-scale datasets, demonstrating the positive impact of model effectiveness and data size on its performance.

\subsection{Short Malicious URLs Study}
\begin{table}[t]
\centering
\small
\caption{Evaluation of baselines and our model under adversarial attack (Training Size = 40000).}
\label{tab:adversarial-results-40000}
\renewcommand{\arraystretch}{1.1}
\begin{tabular}{@{}lccccc@{}}
\toprule
& \multicolumn{4}{c}{\textbf{TPR @ FPR Level}} &  \\
\cmidrule(lr){2-5}
\textbf{Method} & 0.0001 & 0.001 & 0.01 & 0.1 & \textbf{AUC} \\
\midrule
Dephides       & 0.0047 & 0.0651 & 0.2403 & 0.6195 & 0.7221 \\
CGAN & 0.1161 & 0.6617 & 0.7960 & 0.8109 & 0.8854 \\
URLBERT        & 0.1551 & 0.3124 & 0.6170 & 0.8359 & 0.9704 \\
LLaMA-2        & 0.1445 & 0.5920 & 0.7641 & 0.9002 & 0.9672 \\
\midrule
\textbf{Our}   & \textbf{0.1679} & \textbf{0.4823} & \textbf{0.7930} & \textbf{0.9563} & \textbf{0.9896} \\
\bottomrule
\end{tabular}
\end{table}

\begin{table}[t]
\centering
\small
\caption{Evaluation of our model in Short Malicious URLs Study.}
\label{tab:Short}
\setlength{\tabcolsep}{3.3pt} % 控制列间距
\renewcommand{\arraystretch}{1.1} % 控制行间距
\begin{tabular}{lccccc}
\toprule
\textbf{Method} & \textbf{Accuracy} & \textbf{Precision} & \textbf{Recall} & \textbf{F1-score} & \textbf{AUC} \\
\midrule
Dephides       & 0.6000 & 0.7814 & 0.6007 & 0.6792 & 0.8327 \\
CGAN & 0.6799 & 0.8254 & 0.7398 & 0.7802 & 0.9126 \\
URLBERT        & 0.8057 & 0.8875 & 0.7660 & 0.8222 & 0.9390 \\
LLaMA-2        & 0.7985 & 0.8726 & 0.8312 & 0.8513 & 0.9402 \\
\midrule
\textbf{Our}   & \textbf{0.8920} & \textbf{0.9014} & \textbf{0.8250} & \textbf{0.8615} & \textbf{0.9536} \\
\bottomrule
\end{tabular}
\end{table}

Although many detection methods perform well on long URLs, prior studies have shown that short malicious URLs are often more challenging to identify due to their limited lexical and semantic cues, as well as the widespread use of URL shortening services by attackers \cite{maggi2013two,ghalechyan2024phishing}. Figure \ref{fig:url_len} illustrates our analysis of benign and malicious URL lengths in the Grambedding and Mendeley datasets. The results indicate that, although malicious URLs are generally longer on average, a substantial portion of them exhibit lengths comparable to benign URLs. To further validate the robustness of our approach in this challenging scenario, we conduct a dedicated short malicious URL study.

In this experiment, we extract malicious URLs with lengths shorter than 40 characters from the \textit{Grambedding\_dataset}. We randomly select 12,000 samples for training and 10,000 samples for evaluation. The results are summarized in Table~\ref{tab:Short}. As shown, our method consistently outperforms all baselines across metrics, achieving an accuracy of 0.8920 and an AUC of 0.9536. In particular, our approach demonstrates significant improvements in precision (0.9014), recall (0.8250), and F1-score (0.8615), indicating its ability to maintain a favorable balance between identifying true malicious samples and reducing false positives.

Compared to competitive baselines such as URLBERT and LLaMA-2, which also achieve relatively strong performance (AUC around 0.94), our model maintains a clear advantage by delivering both higher accuracy and more balanced precision–recall trade-offs. Meanwhile, traditional models like Dephides and CGAN suffer substantial drops in recall and AUC, suggesting their limited capability in handling short, highly obfuscated URLs. These results confirm that our method is not only effective for detecting long, feature-rich URLs but also robust against short malicious URLs, which represent a more stealthy and security-critical threat in real-world applications.

\section{Discussion}
In this section, we discuss the experimental findings of URL2Graph++ in depth, highlighting its effectiveness, generalization ability, architectural advantages, and practical implications. The analysis synthesizes evidence from data scaling, cross-dataset validation, ablation studies, and adversarial evaluations to provide a comprehensive understanding of the model’s performance and contributions.  

\subsection{Evaluation Metrics}
The results across both balanced (GramBeddings, Kaggle) and imbalanced (Mendeley) datasets consistently demonstrate the superior effectiveness of URL2Graph++ relative to existing baselines. In data scaling experiments, the model maintained clear advantages regardless of training size: even with limited samples (6,400), URL2Graph++ achieved strong detection accuracy and AUC, while at larger scales (32,000 or more) it outperformed advanced baselines such as URLBERT and LLaMA-2 by margins of 1--2\% in accuracy and F1, and by over 1\% in AUC. Similarly, comparative studies against diverse baselines---including Dephides, CGAN, and Transformer-based approaches---consistently confirmed that URL2Graph++ delivers higher precision, recall, and balanced F1 scores. This improvement is attributable to its ability to exploit complementary semantic, structural, and character-level information rather than relying on sequential features alone.  

\subsection{Generalization and Robustness}
A key strength of URL2Graph++ lies in its ability to generalize across heterogeneous data distributions and remain robust under challenging conditions. Cross-dataset evaluations reveal that models trained on GramBeddings and tested on Kaggle, or vice versa, retain high recall and AUC, outperforming baselines that suffered from domain shift. Under extreme imbalance (Mendeley dataset, ratio 1:43), URL2Graph++ maintained stable trade-offs between precision and recall, indicating resilience to skewed distributions. Furthermore, the model demonstrated robustness against adversarial perturbations, achieving consistently higher true positive rates (TPR) at low false positive rate (FPR) thresholds, a critical property for security-sensitive deployments. Its performance on short malicious URLs---often obfuscated and lacking lexical cues---further confirmed its robustness, where URL2Graph++ exceeded strong baselines by over 5\% in F1 and AUC. Collectively, these results highlight that the model achieves not only high in-distribution accuracy but also strong adaptability to domain shifts, imbalanced conditions, adversarial manipulations, and minimal lexical content.  

\subsection{Architectural Advantages}
The architectural design of URL2Graph++ accounts for much of its observed performance. First, the dual-feature encoding integrates semantic subword embeddings from BERT with morphological cues from a character-level CNN, enabling the model to capture both high-level semantics and fine-grained anomalies such as homoglyph substitutions. Second, the introduction of dual-granularity graphs represents a departure from purely sequential URL modeling, providing a principled means of capturing cross-token dependencies at both subword and character levels. This is the first approach to explicitly model URLs as graphs at multiple granularities, aligning with the structural nature of malicious manipulations. Third, the ablation study confirmed that each module contributes distinctly to performance. While the graph-only baseline yielded performance close to random chance, progressively adding semantic encoding (BERT), character-level patterns (CharCNN), and adaptive fusion (DynMM) led to steady gains across all metrics. Notably, simple concatenation of graph and semantic features diluted effective signals, but the proposed DynMM enabled adaptive integration, yielding substantial improvements in F1 and AUC. These findings demonstrate that the strength of URL2Graph++ arises from the complementary and carefully integrated design of its components rather than any single module.  

\subsection{Practical Implications}
The practical implications of these results are twofold. First, the demonstrated scalability of URL2Graph++ confirms its suitability for real-world deployments where large volumes of URL data must be processed efficiently. The ability to sustain high accuracy under low false-positive constraints is particularly critical for automated security pipelines, where false alarms carry operational costs. Second, the adaptability of the model---achieved through dynamic multi-modal fusion---ensures robustness across different types of malicious URLs, ranging from semantically rich phishing attempts to obfuscated short links. While further work is needed to optimize computational efficiency and extend the model to related domains such as domain-name or webpage-level detection, the present results underscore the promise of URL2Graph++ as a practical defense mechanism against evolving online threats.  

\section{Limitations and Future Work}
While URL2Graph++ demonstrates impressive performance across diverse datasets and scenarios, several limitations remain and point to directions for future research:

\begin{itemize}
    \item \textbf{Robustness to Adversarial Manipulation}: Although our method shows resilience under obfuscation and distribution shifts, its robustness against adaptive adversarial attacks has not been fully explored. Future work should investigate potential vulnerabilities and incorporate defense mechanisms such as adversarial training, certified robustness, or anomaly-aware detection strategies.
    
    \item \textbf{Generality Across Domains}: The current design is tailored for malicious URL detection. Its applicability to related domains—such as domain name analysis, webpage content classification, or cross-modal phishing detection—remains to be validated. Exploring transfer learning and domain adaptation techniques could help extend the framework to these settings with minimal architectural modifications.
    
    \item \textbf{Efficiency and Scalability}: While our framework is scalable in mini-batch settings, training and inference costs may become significant for extremely large-scale deployments. Future research may focus on model compression, lightweight graph encoders, or efficient graph construction pipelines to improve deployment feasibility.
\end{itemize}

\section{Conclusion}
In this work, we presented URL2Graph++, a unified malicious URL detection framework that jointly models semantic, structural, and character-level information through dual-granularity graph learning and dynamic fusion. Our method was designed to address two key challenges: the difficulty of generalizing across highly diverse URL distributions, and the resilience required against sophisticated obfuscation strategies employed by attackers. By constructing URL graphs at both subword and character levels, enriching them with semantic embeddings from BERT, and integrating features through a gated fusion mechanism, URL2Graph++ effectively captures contextual semantics, fine-grained anomalies, and cross-token dependencies. Extensive experiments on multiple real-world datasets demonstrated that our approach consistently outperforms state-of-the-art baselines, including Transformer-based and large language models. In particular, URL2Graph++ achieved significant improvements in accuracy, F1-score, and AUC under scenarios of data imbalance, cross-dataset shifts, adversarial perturbations, and short obfuscated URLs. These results confirm the robustness and adaptability of our method in diverse and security-critical environments.

\bibliographystyle{IEEEtran} 
\bibliography{references}

\end{document}